\documentclass[12pt,letterpaper]{article}
\pdfoutput=1

\usepackage[dvips]{epsfig}
\usepackage{color}
\definecolor{refkey}{rgb}{1,0,0}
\definecolor{labelkey}{rgb}{0,0,1}
\usepackage{amsfonts,amssymb,amsmath}
\usepackage{MnSymbol,wasysym}
\usepackage{esint}
\usepackage{float}

\numberwithin{equation}{section}

\newcommand{\be}{\begin{equation}}
\newcommand{\ee}{\end{equation}}
\newcommand{\ben}{\begin{displaymath}}
\newcommand{\een}{\end{displaymath}}
\newcommand{\bea}{\begin{eqnarray}}
\newcommand{\eea}{\end{eqnarray}}
\newcommand{\bean}{\begin{eqnarray*}}
\newcommand{\eean}{\end{eqnarray*}}

\def\s {\sigma}

\newcommand{\ads}[1]{\mbox{${AdS}_{#1}$}}

\newcommand{\ie}{{\it i.e.}}

\newcommand{\commentout}[1]{}

\newcommand{\beq}{\begin{equation}}
\newcommand{\eeq}{\end{equation}}
\newcommand{\beqr}{\begin{displaymath}}
\newcommand{\eeqr}{\end{displaymath}}
\newcommand{\beqa}{\begin{eqnarray}}
\newcommand{\eeqa}{\end{eqnarray}}
\newcommand{\beqar}{\begin{eqnarray*}}
\newcommand{\eeqar}{\end{eqnarray*}}
\newcommand{\cN}{{\cal N}}

\newcommand{\cO}{{\cal O}}

\newcommand{\non}{\nonumber}

\newcommand{\cA}{{\cal A}}

\newcommand{\half}{\ensuremath{\frac{1}{2}}}

\newcommand{\bz}{\ensuremath{\bar{z}}}

\newcommand{\N}[1]{\ensuremath{\cN=#1}}

\renewcommand{\Re}{\ensuremath{\mathrm{Re}}}
\renewcommand{\Im}{\ensuremath{\mathrm{Im}}}

\begin{document}

\title{\LARGE \bf Minimal area surfaces in \ads{n+1} and Wilson loops }

\author{
	 Yifei He\thanks{E-mail: \texttt{he163@purdue.edu}},
	 Changyu Huang\thanks{E-mail: \texttt{cyhuang@purdue.edu}},
	 Martin Kruczenski\thanks{E-mail: \texttt{markru@purdue.edu}} \\
	Department of Physics and Astronomy, Purdue University,  \\
	525 Northwestern Avenue, W. Lafayette, IN 47907-2036.}

\maketitle

\begin{abstract}
 The AdS/CFT correspondence relates the expectation value of Wilson loops in \N{4} SYM to the area of minimal surfaces in \ads{5}.
 In this paper we consider minimal area surfaces in generic Euclidean \ads{n+1} using the Pohlmeyer reduction in a similar way as we did previously in Euclidean \ads{3}. As in that case, the main obstacle is to find the correct parameterization of the curve in terms of a conformal parameter. Once that is done,  the boundary conditions for the Pohlmeyer fields are obtained in terms of conformal invariants of the curve.  After solving the Pohlmeyer equations, the area can be expressed as a boundary integral involving a generalization of the conformal arc-length, curvature and torsion of the curve. Furthermore, one can introduce the $\lambda$-deformation symmetry of the contours by a simple   change in the conformal invariants. This determines the $\lambda$-deformed contours in terms of the solution of a boundary linear problem.
 In fact the condition that all $\lambda$ deformed contours are periodic can be used as an alternative to solving the Pohlmeyer equations and is  equivalent to imposing the vanishing of an infinite set of conserved charges derived from integrability. 
\end{abstract}

\clearpage
\newpage



\section{Introduction}
\label{intro}
 One of the most important results of the AdS/CFT correspondence \cite{Maldacena:1997re,Gubser:1998bc,Witten:1998qj} is the duality between the expectation value of Wilson loops in \N{4} SYM and the area of minimal surfaces in \ads{5} \cite{Maldacena:1998im,Rey:1998ik}. 
 There is a large amount of work on the subject, see for example \cite{WLref} and, in particular \cite{cWL} for the circular Wilson loop, the most studied case. The main interest of this problem is its integrability properties \cite{WLint}.  The basic idea is that 
 the computation of the Wilson loops in the strong coupling limit is translated into finding the area of the minimal surface ending on a boundary curve defined by the Wilson loop. In order to determine the minimal surface dual to a given Wilson loop and find the area, it is important to exploit the integrability of the string sigma model and the conformal invariance of the boundary theory. Recently in \cite{WLMA}, a integrability-based and manifestly conformally invariant formalism was proposed for studying Wilson loops dual to minimal surfaces in Euclidean $AdS_3$. In this case the boundary is $\mathbb{R}^2$ and the curve is given by a function $X(s)=X_1(s)+i X_2(s)$. One needs to find the reparametrization between the conformal angle $\theta$ of the string worldsheet and the arbitrary parameter $s$. Then the Schwarzian derivative of $X(\theta)$ provides the boundary conditions for the Pohlmeyer functions $\alpha$, $f$, and it defines the potential of a Schr\"odinger-like equation whose solutions encode the shape of the curve. In this context, the one-parameter family of boundary curves  expected from integrability \cite{IKZ} can be simply obtained by solving the Schr\"odinger-like equation with the $\lambda$-deformed potential without finding the corresponding minimal surfaces. In \cite{Dekel}, this formalism was applied to study Wilson loops perturbatively away from the circular contour and the area of the dual minimal surfaces was found to high orders.   In \cite{Huang:2016atz}, solutions to the Schr\"odinger-like equation were obtained in terms of Mathieu functions. In \cite{He:2017zsk}, a numerical method was implemented to find the reparametrization $s(\theta)$ for Wilson loops of arbitrary shapes. These provide checks for the applicability of the integrability-based method. For the case of general \ads{n+1} that we consider here, in \cite{Klose} the role of $\lambda$-deformations or master symmetry as generator of the integrable charges was explained.  
 
 In this paper, we generalize the aforementioned \ads{3} formalism to higher dimensional \ads{n+1}. Since the equations for minimal surfaces in \ads{n+1} are also integrable the ideas are similar although the calculations are more involved. In \ads{3} a central role was played by the Schwarzian derivative since it is conformally invariant. For that reason, as a first step, in section \ref{boundarycurve} we introduce an analogous set of conformal invariant quantities $(\zeta,\mu_a)$ associated with a given curve in $\mathbb{R}^n$. They determine the boundary conditions for the Pohlmeyer reduction, transform simply under $\lambda$-deformations, and can be used to define other useful invariants such as the conformal curvature and torsion.  We also define the boundary linear problem which can be solved to reconstruct the curve starting from the invariants. In section \ref{Pohlmeyer}, we review the standard Pohlmeyer reduction to study minimal surfaces in Euclidean $AdS_{n+1}$. In the next section, we show how the boundary conditions for the Pohlmeyer reduction are determined by the conformal invariants on the boundary curve. In section \ref{lambdadeform}, we study the transformation properties of the boundary conformal invariants under $\lambda$-deformations. Here we describe how to construct the $\lambda$-deformed curves from the original one and define conformal and reparametrization invariant quantities which transform simply under $\lambda$-deformation. These quantities are used in section \ref{computearea} to give a formula for the regularized area in the form of a boundary integral. In section \ref{wavyline}, we apply this method on the wavy Wilson line as an example. The last section is a summary.

\section{Conformal invariants for curves in $\mathbb{R}^n$ and a boundary linear problem}\label{boundarycurve}

The construction of conformal invariants for a curve in $\mathbb{R}^n$ can be done in a manner that parallels the Pohlmeyer reduction which we review in the next section. Doing it in this way will ensure a direct relation between the boundary values of the Pohlmeyer functions and the conformal invariants.  

Consider a curve $\mathbf{x}(s)$ in $\mathbb{R}^n$. Define the unit tangent as
\beq
 \mathbf{v} = \frac{\mathbf{x}'}{|\mathbf{x}'|}
\eeq
where we denoted $s$ derivatives with a prime ($\mathbf{x}' = \partial_s \mathbf{x}(s)$).
We also need to define a normal frame, that is, a set of normals $\mathbf{n}_a(s)$ such that
\beq
 \mathbf{n}_a \cdot \mathbf{v}=0, \ \ \ \mathbf{n}_a\cdot \mathbf{n}_b = \delta_{ab}.
\eeq
At each point of the curve there is an $SO(n-1)$ ambiguity in choosing the normals. We introduce the "gauge fields"
\beq\label{boundarygaugefield}
b_{ab}(s) = \mathbf{n}_b\cdot \partial_s \mathbf{n}_a.
\eeq
Since we are interested in studying conformal properties of the curve, we embed $\mathbb{R}^n$ in $\mathbb{R}^{(n+1,1)}$ in such a way that 
the conformal $SO(n+1,1)$ group of $\mathbb{R}^n$ acts linearly in $\mathbb{R}^{(n+1,1)}$. Define now the following $\mathbb{R}^{(n+1,1)}$ basis vectors at each point of the curve:
\beqa
y &=& \frac{1}{|\mathbf{x}'|}(\mathbf{x},\half(\mathbf{x}^2-1) ,\half(\mathbf{x}^2+1) ),\\
y' &=& -\frac{|\mathbf{x}'|'}{|\mathbf{x}'|} y + (\mathbf{v},\mathbf{v}\cdot \mathbf{x},\mathbf{v}\cdot \mathbf{x}), \\
y'' &=& -\left(\frac{|\mathbf{x}'|'}{|\mathbf{x}'|} \right)' y - \frac{|\mathbf{x}'|'}{|\mathbf{x}'|} y' + (\mathbf{v}',\mathbf{v}'\cdot \mathbf{x},\mathbf{v}'\cdot \mathbf{x}) + |\mathbf{x}'| (0,1,1), \\
\label{Nboundary}
n_a &=& (\mathbf{n}_a, \mathbf{n}_a\cdot \mathbf{x},\mathbf{n}_a\cdot \mathbf{x}) + (\mathbf{n}_a\cdot \mathbf{v}') y.
\eeqa
These vectors satisfy:
\begin{equation}
\begin{aligned}
y^2&= 0,\ y'{}^2=1,\ y\cdot y'= 0,\\
y'\cdot y''&= 0,\  y\cdot y''=-1,\\
 y\cdot n_a &= 0,\ y'\cdot n_a = 0,\ y''\cdot n_a=0. 
\end{aligned}
\end{equation}

We can now define the following $SO(n+1,1)$ invariant quantities 
\beqa
\zeta = y''^2 = \mathbf{v}'{}^2 + 2\frac{|\mathbf{x}'|''}{|\mathbf{x}'|}-3\frac{|\mathbf{x}'|'{}^2}{|\mathbf{x}'|^2}, \label{zetadef} \\ 
\mu_a = y'''\cdot n_a = \mathbf{n}_a\cdot(\mathbf{v}''-\frac{|\mathbf{x}'|'}{|\mathbf{x}'|}\mathbf{v}'). \label{mudef}
\eeqa
We should note that the $\mu_a$'s are not gauge invariant; namely they depend on the choice of normals and therefore they are conformally invariant only up to gauge transformations. Further, $\zeta$ and $\mu_a$ are not reparameterization invariant. Actually, under a
change of parameter $s\rightarrow \sigma$, they transform as
\beqa\label{zetatrans}
\zeta(\sigma) &=&  (\partial_\sigma s)^2 \zeta(s) + 2\{s,\sigma\}, \\ \label{mutrans}
\mu_{a}(\sigma) &=& (\partial_\sigma s)^2 \mu_{a}(s).
\eeqa 
Namely, $\zeta(\s)$ and $\mu_a(\s)$ transform like the real and imaginary parts of the Schwarzian derivative in \ads{3}, thus providing a natural higher dimensional generalization. For that reason, we choose $\zeta(\s)$ and $\mu_a(\s)$ as the quantities to characterize the curve in a conformally invariant way. Later it will be useful to define quantities that are also gauge  invariant, for example 
\beq
 \mu =\sqrt{ \sum_a \mu_a^2}, \ \ \ T(s)=\frac{|D_s \hat{\mu}_a|}{\sqrt{\mu(s)}},  
\eeq
where $T(s)$ is called the conformal torsion (see \cite{ConformalCurve} for the more standard definition that agrees with the present one\footnote{See appendix \ref{appA} for a detailed comparison.}), $\hat{\mu}_a=\frac{\mu_a}{\mu}$, and the covariant derivative is defined as
\beq
 D_s \mu_a = \partial_s \mu_a - b_{ab}\, \mu_b.
\eeq
The torsion $T(s)$ is not only conformally invariant and gauge invariant but also reparameterization invariant. In fact, $\mu$ can be used to define a conformal arc-length one-form $\omega=\sqrt{\mu} ds$ which is invariant under $s\to\s$ and allows to introduce the reparameterization invariant function
\beq
 \chi(s) = \int_A^s \sqrt{\mu(s')} ds'
\eeq
 Here $A$ is an arbitrary point on the curve. With $\chi$, we can define another reparameterization invariant quantity, known as the conformal curvature $Q$ \cite{ConformalCurve}:
\beq
Q(s)=-\frac{1}{2}\frac{\zeta(s)-2\{\chi,s\}}{\mu(s)}.
\eeq
where $\{\chi,s\}=\frac{\chi'''}{\chi'}-\frac{3}{2}\left(\frac{\chi''}{\chi'}\right)^2$ denotes the Schwarzian derivative.
The $\chi$, $Q$ and $T$ are conformally as well as gauge and reparametrization invariant quantities of the curve and will play an important role in computing the area. For some purposes it is natural to parameterize the curve using $\chi(s)$ in which case $\hat{\mu}_a(\chi)=\mu_a(\chi)$ and the conformal curvature and torsion become simply
\beq
  Q(\chi) = -\half\zeta(\chi), \ \ \ T(\chi) = |D_\chi \mu_a(\chi)|
\eeq 
One can define further invariants by taking higher order derivatives but these are the only invariants needed in the rest of the paper.

Going back to the gauge dependent quantities, it is useful to note that
\beqa
 y''' &=& - \zeta y' -\half \zeta' y + \mu_a n_a, \\
 n'_a &=& b_{ab} n_b + \mu_a y.
\eeqa
We can summarize the derivatives as
\beq
 \partial_s\left(\begin{array}{c}y\\ y'\\ y'' \\ n_a\end{array}\right) =
\left(
 \begin{array}{cccc} 0&1&0&0 \\ 
                                 0&0&1&0 \\
                                 -\half\zeta'&-\zeta&0&\mu_b \\
                                 \mu_a&0&0&b_{ab}
\end{array}\right)
 	\left(\begin{array}{c}y\\ y'\\ y'' \\ n_b\end{array}\right) 
\eeq
From here it is clear that, if $\zeta$, $\mu_a$ and $b_{ab}$ are given  we can reconstruct the curve, up to a conformal transformation, by solving the linear problem
\beq
 \partial_s\left(\begin{array}{c}\alpha_1\\ \alpha_2\\ \alpha_3 \\ \beta_a\end{array}\right) =
 \left(
 \begin{array}{cccc} 0&1&0&0 \\ 
 	0&0&1&0 \\
 	-\half\zeta'&-\zeta&0&\mu_b \\
 	\mu_a&0&0&b_{ab}
 \end{array}\right)
 \left(\begin{array}{c}\alpha_1\\ \alpha_2\\ \alpha_3\\ \beta_b\end{array}\right)  
 \label{blp}
\eeq
or equivalently
\beqa
 \alpha_1'''+\zeta \alpha_1'+\half \zeta' \alpha_1 &=& \mu_a\beta_a, \\
 D_s \beta_a &=& \mu_a \alpha_1.
\eeqa
It is necessary to find $n+2$ linearly independent solutions $\alpha_1^{\mu=1\ldots n+2}$ that can be assembled into the vector $y^\mu=\alpha_1^\mu$. Later we will see that the $\lambda$ deformation symmetry can be introduced as a change in the functions $\mu_a$. 
After that, the linear problem just described can be used to reconstruct the $\lambda$-deformed curves. 

\section{Pohlmeyer reduction}\label{Pohlmeyer}

 The standard approach to study minimal surfaces in Euclidean $AdS_{n+1}$ is through the Pohlmeyer reduction \cite{Pohlmeyer}. Here we use the notation of \cite{Hoare}. The string action is given by
 \begin{equation}\label{action}
 S=\frac{1}{2}\int d\sigma d\tau(\partial Y\cdot\bar{\partial}Y)-\Lambda(Y\cdot Y+1).
 \end{equation}
 The equations of motion are
\beq\label{EoM}
 \partial\bar{\partial} Y^\mu = \Lambda Y^\mu,
 \eeq
 where $\Lambda$ is a Lagrange multiplier imposing the constraint  $Y\cdot Y=-1$ and is given by
 \begin{equation}
 \Lambda=\partial Y\cdot \bar{\partial}Y.
 \end{equation}
 It should be supplemented by the Virasoro constraints
\beq\label{Virasoro}
 \partial Y\cdot \partial Y = 0 = \bar{\partial} Y\cdot \bar{\partial} Y .
\eeq
Now we introduce extra vectors $N^\mu_a$ with $a=1,...,n-1$ so that we have a basis
\beq
\{ \partial Y^\mu, \bar{\partial}Y^\mu,N^\mu_a,Y^\mu\}
\eeq
with $N_a\cdot Y=0$, $N_a\cdot \partial Y=0$, $N_a\cdot\bar{\partial}Y=0$, and $N_a\cdot N_b=\delta_{ab}$.
We also define the functions $\alpha$, $u_a$ and $B_{ab}$ as:
\begin{equation}
\label{lambdaalpha}
\Lambda=\frac{1}{2}e^{2\alpha},
\end{equation}
\begin{equation}
\label{uubar}
u_a= 2 N_a\cdot \partial^2 Y, \quad \bar{u}_a=2N_a\cdot \bar{\partial}^2 Y,
\end{equation}
\begin{equation}
\label{BBbar}
B_{ab}= N_b\cdot \partial N_a, \quad \bar{B}_{ab}= N_b\cdot \bar{\partial} N_a.
\end{equation}
Using the equations of motion and the constraints we find that 
\beqa
 \partial^2 Y^\mu &=& 2\partial \alpha \partial Y^\mu + \frac{1}{2} u_a N_a^\mu, \\
 \bar{\partial}^2 Y^\mu &=& 2\bar{\partial} \alpha \bar{\partial} Y^\mu +\frac{1}{2} \bar{u}_a N_a^\mu, \\
 D N_a^\mu &=&  -u_a e^{-2 \alpha} \bar{\partial} Y^\mu, \\
 \bar{D} N_a^\mu &=& -\bar{u}_a e^{-2 \alpha} \partial Y^\mu, 
\eeqa
where the covariant derivatives are
\beq
 D v_a = \partial v_a - B_{ab} v_b, \ \ \bar{D} v_a = \bar{\partial} v_a - \bar{B}_{ab} v_b.
\eeq
Using the commutators $[\partial,\bar{\partial}]=0$ and $[D,\bar{D}]=F_B$ where $F_B$ is the field strength associated with the gauge field $B_{ab}$, we obtain the following Pohlmeyer equations: 
\beqa
\label{coshgordon}
\partial\bar{\partial} \alpha &=& \frac{1}{4}(e^{2\alpha}+u_a\bar{u}_a e^{-2\alpha}), \\
\bar{D} u_a &=& \bar{\partial} u_a - \bar{B}_{ab} u_b =0, \\
D \bar{u}_a &=& \partial \bar{u}_a - B_{ab} \bar{u}_b =0, \\
F_B &=& \bar{\partial} B_{ab} - \partial \bar{B}_{ab} + B_{ac}\bar{B}_{cb} - \bar{B}_{ac}B_{cb} =\frac{1}{2} (u_a\bar{u}_b-\bar{u}_a u_b)e^{-2\alpha}.
\eeqa

It is also convenient to use a normalized basis $\{e^{-\alpha}\partial_\s Y^\mu, e^{-\alpha}\partial_\tau Y^\mu,N^\mu_a,Y^\mu\}$ and write
\beq
 \partial\left(\begin{array}{c} e^{-\alpha}\partial_\s Y^\mu \\ e^{-\alpha}\partial_\tau Y^\mu \\ N_a^{\mu} \\ Y^{\mu}\end{array}\right) =
V \left(\begin{array}{c} e^{-\alpha}\partial_\s Y^\mu \\ e^{-\alpha}\partial_\tau Y^{\mu}\\ N_a^{\mu} \\ Y^{\mu}\end{array}\right)
\eeq
and its complex conjugate. Here $V$ can be written as
\beq
 V = \frac{1}{2} e^{\alpha} (M_{0n}+iM_{0n+1}) + i \partial\alpha M_{n n+1} -\frac{1}{2} e^{-\alpha} u_a(M_{an}-i M_{an+1}) + \half B_{ab}M_{ab}
\eeq
where we labeled the rows and columns of $V$ as $(a=1,...,n-1,n,n+1,0)$. The matrices $M_{\mu\nu}$ are the generators of $SO(n+1,1)$ in the vector representation (see appendix). The consistency condition for this linear system
implies that
\beq
\bar{\partial} V - \partial \bar{V}+ V\bar{V}- \bar{V} V =0
\eeq
namely the Pohlmeyer current
\beq
 j_P = V dz + \bar{V} d\bz
\eeq
is flat. This is a direct consequence of the Pohlmeyer equations and the commutation relations of the $M_{\mu\nu}$ matrices and therefore is independent of the representation used for them. Sometimes it is convenient to write the matrices $M_{\mu\nu}$ in another representation, for example the spinor one but we will not do so in this paper. 
Although the Pohlmeyer reduction provides a well-known way to solve the minimal surface equations, we have to find the particular solution associated with a given curve,  namely, we have to find the boundary values of the Pohlmeyer fields in terms of the shape of the boundary curve. 
 In the next section, we find that those boundary values are related to the conformal invariants of the boundary curve described in section \ref{boundarycurve}.

\section{Boundary conditions for the Pohlmeyer reduction}\label{Pohlmeyerbc}

To determine the boundary conditions for the Pohlmeyer reduction, we perform an expansion near the boundary. In appendix \ref{appB}, following \cite{Polyakov:2000jg} we describe an expansion in Poincar\'{e} coordinates that can be easily converted in an expansion in embedding coordinates as needed here.  Alternatively, we can solve the equations directly in embedding coordinates obtaining the same result.

More precisely, the world-sheet is taken to be the upper half plane $z=(\sigma,\tau>0)$, and one approaches the boundary by taking $\tau\to0$. Using the equation of motion \eqref{EoM} and Virasoro constraints \eqref{Virasoro}, the embedding coordinates and the Lagrange multiplier $\Lambda$ have expansions in $\tau$ of the following form:
\beq
 Y^\mu(\sigma,\tau) = \frac{1}{\tau} y^\mu_0(\sigma) +  (\half y''^\mu_0(\sigma) -2 \Lambda_2(\sigma) y^\mu_0(\sigma))\tau +  y_3^\mu(\sigma)\tau^2 + \cO(\tau^3),
\eeq
\beq\label{Lambdaexpand}
 \Lambda(\sigma,\tau) = \frac{1}{2\tau^2} + \Lambda_2(\sigma) + \Lambda_4(\sigma) \tau^2 +\cO(\tau^3).
\eeq
Here $y_0^{\mu}(\sigma)$ is the boundary curve in embedding coordinates, and $\Lambda_2(\sigma)$ is given by 
\beq
 \Lambda_2(\sigma) = -\frac{1}{6} y''^2_0(\sigma).
\eeq
Meanwhile, the following relations hold:
\beq
 y_0^2 = 0, \ \ y'_0{}^2=1, \ \ y_0\cdot y_{3}=0\ \ y'_0\cdot y_{3} =0,
\eeq
where prime denotes taking derivative with respect to $\sigma$.
Using the relation \eqref{lambdaalpha}, it is easy to see that
\begin{equation}
(\partial^2\alpha-(\partial\alpha)^2)|_{\tau\to0}=\frac{1}{2}(\partial^2\ln \Lambda-\frac{1}{2}(\partial\ln\Lambda)^2)|_{\tau\to0}=-\frac{3}{2}\Lambda_2.
\end{equation}
One can also expand the $N_a$'s in terms of $\tau$:
\begin{equation}
N_a^\mu(\sigma,\tau)=n_{a0}^{\mu}(\sigma)+n_{a1}^{\mu}(\sigma)\tau+n_{a2}^{\mu}(\sigma)\tau^2+n_{a3}^{\mu}\tau^3+\cO(\tau^4).
\end{equation}
From the orthogonality of $N_a$ with $Y$, $\partial Y$ and $\bar{\partial}Y$, one has
\begin{equation}
\begin{aligned}
y_0\cdot n_{a0}=0,&\quad y'_0\cdot n_{a0}=0,\quad y''_0\cdot n_{a0}=0,\\
y_0\cdot n_{a1}=0,&\quad y'_0\cdot n_{a1}=0, \quad y''_0\cdot n_{a1}=-3y_3\cdot n_{a0},\\
y_0\cdot n_{a2}=0,&\quad y'_0\cdot n_{a2}=-\frac{1}{2}y'''_0\cdot n_{a0},\quad y_0\cdot n_{a3}=\frac{1}{2}y_3\cdot n_{a0}.
\end{aligned}
\end{equation}
Plugging the series expansion of $Y^{\mu}$ into the definitions \eqref{uubar}, we obtain the boundary values of $u_a$ and $\bar{u}_a$:
\begin{equation}
\begin{aligned}
u_a(\sigma,\tau=0)&= - \nu_a(\s) - i \mu_a(\s),\\
\bar{u}_a(\sigma,\tau=0)&= - \nu_a(\s) + i \mu_a(\s),
\end{aligned}
\end{equation}
with the definitions
\beqa
 \nu_a(\s) &=& 3y_3(\s)\cdot n_{a0}(\s), \label{nudef} \\
 \mu_a(\s) &=& y'''_0(\s)\cdot n_{a0}(\s).  \label{mudef1}
\eeqa
Here $\mu_a(\s)$ is the same as the one defined in \eqref{mudef} and $\nu_a(\s)$ is a new quantity depending on $y_3(\s)$, or equivalently $X^i_3(\s)$ as defined in \eqref{B6}. 
As can be seen from \cite{Polyakov:2000jg} the value of $\nu_a$ is related to the variation of the area with respect to changes in the shape of the contour (see appendix \ref{appB}) and therefore can be thought as a conjugate momentum to the shape.
Furthermore, from eq. \eqref{BBbar}, we define
\begin{equation}
B_{ab}^{(\s)}=B_{ab}+\bar{B}_{ab}=N_b\cdot\partial_\sigma N_a.
\end{equation}
The boundary value of $B_{ab}^{(\s)}$ is given by
\begin{equation}
b_{ab0}(\sigma)=n_{b0}(\sigma)\cdot\partial_\sigma n_{a0}(\sigma).
\end{equation}

\begin{figure}[!h]
\centering
\includegraphics[trim=0cm 0cm 0cm 0cm, clip=true, width=1.00 \textwidth]{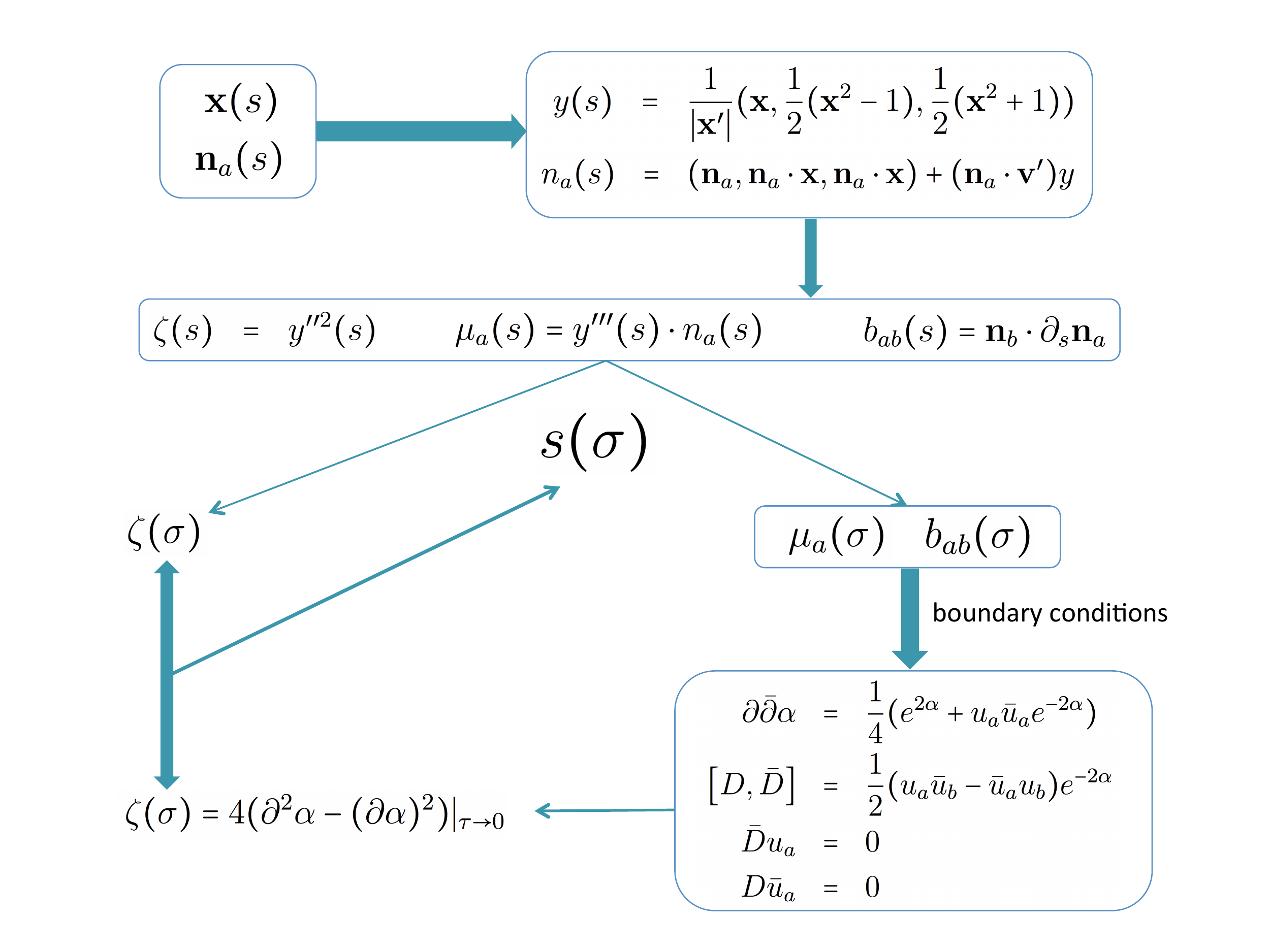}
\caption{The procedure for finding the reparametrization $s(\s)$ and solving the Pohlmeyer equations. Given the curve $\mathbf{x}(s)$ and a set of normals $\mathbf{n}_a(s)$ we compute $(\zeta(s),\mu_a(s), b_{ab}(s))$. Then a reparameterization $s(\s)$ is proposed and $(\zeta(\s),\mu_a(\s), b_{ab}(\s))$ computed. The values of $(\mu_a(\s), b_{ab}(\s))$ are used as boundary values for the Pohlmeyer reduction. Once that is solved, the limiting value of $\alpha$ at the boundary determines $\zeta(\s)$ independently. If it does not agree with the previously computed one, the difference should be minimized by changing $s(\s)$ until they agree.}
\label{procedure}
\end{figure}

Comparing the expression of $\Lambda_2(\sigma)$, $u_a(\sigma)$, and $b_{ab0}(\sigma)$ with the boundary conformal invariants and gauge fields we defined in section \ref{boundarycurve}, we can identify the following limiting values:
\beqa\label{Pb1}
4(\partial^2\alpha-(\partial\alpha)^2)|_{\tau\to0}&=&\zeta(\sigma),\\ \label{Pb2}
-\text{Im}(u_a(\sigma,\tau=0))&=& \mu_a(\sigma),\\ \label{Pb3}
b_{ab0}(\sigma)&=& b_{ab}(\sigma).
\eeqa
In the last equation, we take $n_{a0}$ to be of the form \eqref{Nboundary} and used that
\begin{equation}
n_{b0}\cdot\partial_\sigma n_{a0}=\mathbf{n}_b\cdot \partial_\sigma \mathbf{n}_a+(\mathbf{n}_a\cdot \mathbf{v}')y\cdot(\partial_\sigma \mathbf{n}_a,\partial_\sigma \mathbf{n}_a\cdot \mathbf{x},\partial_\sigma \mathbf{n}_a\cdot \mathbf{x}),
\end{equation}
with the second term vanishing. Equations \eqref{Pb1}, \eqref{Pb2} and \eqref{Pb3} state the relation between the boundary values of the Pohlmeyer functions and the conformal invariants of the curve.  One could use this to solve the Pohlmeyer equations,  find the functions $\zeta$, $\mu_a$ and $b_{ab}$, and reconstruct the boundary curve up to a conformal transformation by solving the boundary linear problem described in section \ref{boundarycurve}. In this paper the idea is the opposite, we want to use these relations as boundary condition for the Pohlmeyer equations, in the same way as done in \cite{WLMA} for \ads{3}. However, as seen before in \eqref{zetatrans} and \eqref{mutrans}, $\zeta(\s)$ and $\mu_a(\s)$,  are not reparameterization invariant, and neither is $b_{ab}$ which transform like
\begin{equation}\label{btrans}
b_{ab}(\s)=(\partial_\s s)b_{ab}(s).
\end{equation}
Therefore, given a curve in terms of an arbitrary parameter $s$, the reparameterization $s(\s)$ has to be found.  The idea is similar to the case of $AdS_3$ \cite{Dekel,He:2017zsk}. One should propose a reparameterization $s(\s)$ and use $\mu_a$ and $b_{ab}$ as boundary conditions to solve the Pohlmeyer equations ($\zeta(\s)$ is not needed for this). After that there are two computations of $\zeta(\s)$, one directly from the boundary curve \eqref{zetadef}, and the other from the limiting value of the conformal factor $\alpha$ \eqref{Pb1}. If both agree, then $s(\s)$ is the correct reparameterization. If they disagree, the difference is a measure of the error that can be minimized numerically.
 This way to define the problem is better illustrated in figure \ref{procedure}.
To summarize this section, we obtained a clear picture, similar to the one in \ads{3} to approach the problem for generic \ads{n+1}. It is not a direct computation of the area but could be converted into one by using a numerical procedure as in \cite{He:2017zsk} or by expanding near the straight line \cite{Dekel}. The main result up to here is the correct identification of the boundary values of the Pohlmeyer field in terms of conformal invariants and a precise test of the conformal reparameterization (or conformal gauge), namely the agreement of two alternative computations of $\zeta(\s)$.  Notice that, as discussed in \cite{WLMA}, this mimics what has been done before for the flat space case. 

\section{$\lambda$-deformations}\label{lambdadeform}

 A manifestation of integrability is that the Pohlmeyer flat current can be deformed with a complex parameter $\lambda$ while remaining flat, a symmetry that can be used to generate an infinite tower of conserved charges \cite{Klose}.  When $|\lambda|=1$, the reality conditions are preserved and we obtain a one parameter family of boundary curves and minimal surfaces with the same area.  In terms of the Pohlmeyer functions, the $\lambda$-deformation is the simple replacement $u_a\rightarrow \lambda u_a$, $\bar{u}_a \rightarrow \frac{1}{\lambda}\bar{u}_a$. The deformed flat current is written as
\begin{equation}
\begin{aligned}
 V &= \frac{1}{2} e^{\alpha} (M_{0n}+iM_{0n+1}) +i \partial \alpha M_{n n+1} -\frac{1}{2} e^{-\alpha} \lambda u_a(M_{an}-i M_{an+1}) + \half B_{ab}M_{ab},\\
  \bar{V} &= \frac{1}{2} e^{\alpha} (M_{0n}-iM_{0n+1}) - i \bar{\partial} \alpha M_{n n+1} -\frac{1}{2} e^{-\alpha} \frac{1}{\lambda} \bar{u}_a(M_{an}+i M_{an+1}) + \half \bar{B}_{ab}M_{ab}.
\end{aligned}
\end{equation}
Again, one can choose a representation for the generators $M_{\mu\nu}$ and solve the deformed linear problem to find the deformed surfaces and boundary curves.

To see the transformation properties of the boundary quantities under $\lambda$-deformation, we look at the boundary conditions of the Pohlmeyer reduction. The functions $\zeta(\s)$ and $b_{ab}(\s)$ are unchanged by the $\lambda$-deformation. Meanwhile, the function $\mu_a(\s)$ should be modified to
\begin{equation}
\mu_a(\s) \rightarrow \mu^{\lambda}_a(\sigma)=-\text{Im}(\lambda u_a(\sigma,\tau=0)),
\end{equation}
i.e., the $\mu^\lambda_a$ in the boundary linear problem \eqref{blp} for the $\lambda$-deformed curve is given by
\begin{equation}\label{lambdamu}
\mu_a^{\lambda}(\sigma)=-\frac{i}{2}(\lambda-\frac{1}{\lambda})\nu_a(\s)+\frac{1}{2}(
\lambda+\frac{1}{\lambda})\mu_a(\s).
\end{equation}
We see that the gauge dependent conformal invariant quantities $\mu_a(\s)$ transform simply under $\lambda$-deformations. To reconstruct the $\lambda$-deformed curve, one needs to solve the boundary linear problem \eqref{blp} with the potentials $\zeta(\s)$, $b_{ab}(\s)$ and $\mu_a^{\lambda}(\s)$. 

Another useful result is that, since the boundary value of $u_a$ has the same properties as $\mu_a$ under conformal transformations and reparameterizations they can be used to define a generalized version of the conformal arc-length, curvature and torsion that play an important role in determining the area. Thus, we define a generalized conformal arc-length:
\beq
\tilde{\chi}(\s) = \int_A^\s \sqrt{U(\s')} d\s',
\eeq
where $U(\s)$ is the boundary value of the holomorphic function
\beq
 U(z) = \sqrt{u_a u_a }.
 \label{Udef}
\eeq
The holomorphicity of $U(z)$ is easily seen by observing $\bar{\partial} (u_a u_a) = 2 u_a \bar{D}u_a=0$.
Under $\lambda$-deformations, $U\to \lambda U$, and
\begin{equation}\label{chilambda}
\tilde{\chi}^{\lambda}(\s)= \sqrt{\lambda} \tilde{\chi}(\s).
\end{equation}
implying that $\{\tilde{\chi},\s\}$ is invariant.
We can then define the generalized conformal curvature and torsion accordingly
\beqa
\tilde{Q}(\s)&=&-\frac{1}{2}\frac{\zeta(\sigma)-2\{\tilde{\chi},\sigma\}}{U(\sigma)},\\
\tilde{T}(\s)&=&\frac{|D_\sigma \hat{u}_a|}{\sqrt{U(\sigma)}},
\eeqa
where $\hat{u}_a=\frac{u_a}{U}$. They also transform simply under $\lambda$-deformations:
\beqa\label{Qlambda}
\tilde{Q}^{\lambda}(\s)&=&\frac{1}{\lambda}\tilde{Q}(\s),\\ \label{Tlambda}
\tilde{T}^{\lambda}(\s)&=&\frac{1}{\sqrt{\lambda}}\tilde{T}(\s).
\eeqa
The quantities $\tilde{\chi}$, $\tilde{Q}$ and $\tilde{T}$ are again gauge independent and invariant under conformal transformation and reparametrization but now they also transform very simply under $\lambda$-deformations. In the next section, we will see that an integral of a $\lambda$-independent combination of $\tilde{\chi}$, $\tilde{Q}$ and $\tilde{T}$ gives the area of the minimal surface.

\section{Computation of the Area}\label{computearea}

In this section, we derive the formula for the regularized area in $AdS_{n+1}$ and give a formula for the area in terms of a boundary integral of the generalized conformal invariants introduced in the previous section. From the string action \eqref{action}, the area of the minimal surface is given by
\begin{equation}\label{infarea}
\mathcal{A}_{\infty}=2\int d\sigma d\tau \Lambda=\int d\sigma d\tau e^{2\alpha}.
\end{equation}
This integral diverges near the boundary and needs to be regularized. The way to find the regularized area is similar to the case of Euclidean $AdS_3$ \cite{WLMA}. 

Take a contour at $Z=\epsilon$ where $Z$ is the Poincar\'{e} coordinate and $\epsilon$ is small. The area can then be expanded in terms of $\epsilon$:
\begin{equation}
\mathcal{A}_{\infty}=\frac{\mathcal{A}_\epsilon}{\epsilon}+\mathcal{A}_{f}+O(\epsilon^2).
\end{equation}
From eq. \eqref{infarea} and \eqref{coshgordon}, we have
\begin{equation}
\mathcal{A}_{\infty}=-\int_{Z=\epsilon}\nabla\alpha\cdot\frac{\nabla Z}{|\nabla Z|}dl-\int d\sigma d\tau u_a\bar{u}_ae^{-2\alpha}.
\end{equation}
In the appendix, we give the boundary expansion of $Z$. It is easy to see that
\beqa
\partial_\tau Z&=&|X_0'|+3\tau^2 Z_3+\cO(\tau^3),\\
\partial_\s Z&=& \tau |X_0'|'+\tau^3 Z_3'+\cO(\tau^3),
\eeqa
and
\begin{equation}
|\nabla Z|=|X_0'|+\tau^2 (3 Z_3+\frac{1}{2}\frac{{|X_0'|'}^2}{|X_0'|})+\cO(\tau^3).
\end{equation}
Meanwhile, from eq. \eqref{lambdaalpha} and \eqref{Lambdaexpand}, we see that near the boundary, $\alpha$ has the expansion
\begin{equation}
\alpha(\sigma,\tau)=-\ln \tau+\cO(\tau^2),
\end{equation}
and
\beqa
\partial_\tau \alpha&=&-\frac{1}{\tau}+\cO(\tau),\\
\partial_\s\alpha&=&\cO(\tau^2).
\eeqa
Therefore, we have
\begin{equation}
\nabla\alpha\cdot\frac{\nabla Z}{|\nabla Z|}=-\frac{1}{\tau}+\cO(\tau).
\end{equation}
Near the boundary, one has $dl=d\s+\cO(\tau^2)$. We see that
\begin{equation}
\begin{aligned}
\mathcal{A}_\infty&=\int\frac{1}{\tau}d\s-\int d\sigma d\tau u_a\bar{u}_ae^{-2\alpha}\\
&=\int\frac{1}{\epsilon}|X_0'|d\s-\int d\sigma d\tau u_a\bar{u}_ae^{-2\alpha}\\
&=\frac{L}{\epsilon}-\int d\sigma d\tau u_a\bar{u}_ae^{-2\alpha}.
\end{aligned}
\end{equation}
In the second equation we use that $Z=\epsilon=\tau |X_0'|$. Therefore, the finite area is given by
\begin{equation}
\mathcal{A}_f=-\int d\sigma d\tau u_a\bar{u}_ae^{-2\alpha}.
\end{equation}

 Using the holomorphic function $U(z)$ defined in the previous section \eqref{Udef}, we can write
\beq
\tilde{\chi}(z) = \int_A^z \sqrt{U(z')} dz',
\eeq
where $A$ is an arbitrary point in the world-sheet that, for convenience, we take to be at the boundary. Using the Pohlmeyer equations, we can prove that the following current is closed
\beqa
 j_z &=& \frac{2}{\sqrt{U}}\left[(\partial\alpha)^2-\partial^2\alpha-\frac{Du_aDu_a}{4U^2}+\frac{\partial^2U}{4U}-\frac{1}{16}\left(\frac{\partial U}{U}\right)^2\right]  \nonumber\\
 &=& \frac{2}{\sqrt{U}}\left[(\partial\alpha)^2-\partial^2\alpha+\frac{1}{2}\{\tilde{\chi},z\}-\frac{1}{4}D\hat{u}_a D\hat{u}_a\right] , \label{jArea} \\
j_{\bz} &=& -\frac{u_a\bar{u}_a}{\sqrt{U}}e^{-2\alpha} \nonumber,
\eeqa
where $\hat{u}_a=\frac{u_a}{U}$. Notice that there are terms in $j_z$ that are purely holomorphic. They were added to make $j$ a one form under holomorphic coordinate transformations:
\beqa
z &\rightarrow& w(z) , \\
\alpha &\rightarrow& \alpha + \half \ln \frac{\partial z}{\partial w} +  \half \ln \frac{\partial \bar{z}}{\partial \bar{w}} ,\\
u_a &\rightarrow & \left(\frac{\partial z}{\partial w}\right)^2 u_a, \ \ \ \  \bar{u}_a  \rightarrow \left(\frac{\partial \bar{z}}{\partial \bar{w}}\right)^2 \bar{u}_a, \\
B_{ab} &\rightarrow& \frac{\partial z}{\partial w} B_{ab} , \ \ \ \ \bar{B}_{ab} \rightarrow \frac{\partial \bar{z}}{\partial \bar{w}} \bar{B}_{ab} ,  \\
j_z &\rightarrow& j_w=\frac{\partial z}{\partial w} j_z , \ \ \ \ j_{\bar{z}} \rightarrow j_{\bar{w}}=\frac{\partial \bar{z}}{\partial \bar{w}} j_{\bar{z}} .
\eeqa 
 We can now write the finite part of the area as
\beq
 \cA_f = - \int d\s d\tau u_a \bar{u}_a e^{-2\alpha} = \frac{i}{2}\int d(\tilde{\chi} j) = \frac{i}{2}\int_{\mathbb{R}} \tilde{\chi} j,
\eeq
where we used $d\s \wedge d\tau=\frac{i}{2} dz \wedge d\bar{z}$. This integral is performed on the boundary of the upper half plane, \ie\ the real line $\mathbb{R}$ parameterized by $\sigma$.  Explicitly, we have
\begin{equation}
\mathcal{A}_f=\frac{i}{2}\int\tilde{\chi}(\sigma)(j_z\partial_\sigma z+j_{\bar{z}}\partial_\sigma \bar{z}) d\sigma.
\end{equation}
On the boundary, $j_{\bar{z}}$ vanishes and $\partial_\sigma z=1$. We have the following boundary values:
\beqa
U(\sigma)^2&=&\nu_a^2-\mu_a^2+2i\nu_a\mu_a,\\
\{\tilde{\chi},z\}|_{\mathbb{R}}&=&\{\tilde{\chi},\sigma\},\\
D\hat{u}_aD\hat{u}_a|_{\mathbb{R}}&=& D_{\sigma}\hat{u}_aD_{\sigma}\hat{u}_a.
\eeqa
where we used $\bar{D}\hat{u}_a=0$ to replace $D\hat{u}_a=D_\sigma\hat{u}_a$. 
The final formula for the area in terms of the boundary data is
\begin{equation}
\begin{aligned}
\mathcal{A}_f&=i\int\frac{\tilde{\chi}(\sigma)}{\partial\tilde{\chi}(\sigma)}(-\frac{1}{4}\zeta(\sigma)+\frac{1}{2}\{\tilde{\chi},\sigma\}-\frac{1}{4}D_{\sigma}\hat{u}_aD_{\sigma}\hat{u}_a)d\sigma\\
&=\frac{i}{4}\int(2\tilde{Q}-\tilde{T}^2)\tilde{\chi}d\tilde{\chi}, \label{areaF}
\end{aligned}
\end{equation}
where $\tilde{\chi}$, $\tilde{Q}$, and $\tilde{T}$ are the generalized conformal arc-length, curvature and torsion. Using eq. \eqref{chilambda}, \eqref{Qlambda} and \eqref{Tlambda}, we see that the integrand in the area formula is $\lambda$-independent as expected from integrability.
Notice from \ref{appA} that the combination $2Q-T^2$ that appears (in generalized form) in the area formula also appears in the standarized expansion of the curve around a point. In the case of \ads{3} the torsion vanishes and the formula reduces to the one found in \cite{WLMA} although now we have a nicer geometric interpretation of the integrand in terms of a generalized conformal curvature. Finally, notice that this formula relies on $U(z)$ having no zeros on the world-sheet, otherwise extra terms given by integrals around the cuts of $\tilde{\chi}$ are required (see \cite{He:2017zsk}).  

\section{Wavy Wilson line}\label{wavyline}
The wavy Wilson line was originally studied by Semenoff and Young in \cite{SY} and greatly improved in \cite{Dekel}, there has been renewed interest in this topic due to the possibility of defining a 1d conformal theory on the Wilson line \cite{Cooke,Giombi}. In this section, we take the wavy line as an example of applying the formalism described in the previous sections.

\subsection{Boundary curve and conformal invariants}
The shape of the wavy line in $\mathbb{R}^n$ is given by
\begin{equation}
\mathbf{x}(s)=(\epsilon\xi_a(s),s),
\end{equation}
and we are interested in a power series expansion for $\epsilon\rightarrow 0$. Here $s$ being the conformal parameter for the straight line but notice that for the wavy line $s$ will no longer be a conformal parameter and a reparameterization $s(\s)$ will have to be found.
The tangent and normals of the straight line are
\beqa
\mathbf{v}^{(0)}&=&(0,...,0,1),\\
\mathbf{n}_a^{(0)}&=&(0,...,1,...,0), 
\eeqa
where $1$ is the $a$th component in the second equation.
It is then straightforward to write a particular set of normals to the wavy line 
\begin{equation}
\mathbf{n}_a=\mathbf{n}_a^{(0)}-\epsilon\xi'_a\mathbf{v}^{(0)}-\frac{1}{2}\epsilon^2\xi'_a\xi'_b\mathbf{n}_b^{(0)}+\frac{1}{2}\epsilon^3\xi'^2_b\xi'_a\mathbf{v}^{(0)}+\cO(\epsilon^4),
\end{equation}
where prime denotes $s$ derivative.
Using the definitions in section \ref{boundarycurve}, one obtains the conformal invariants and the gauge field on the boundary
\beqa
\zeta(s)&=&\epsilon^2 (3\xi''^2_a+2\xi'''_a\xi'_a)+\cO(\epsilon^4),\\
\mu_a(s)&=&\epsilon\xi'''_a-\frac{1}{2}\epsilon^3(\xi'_a\xi'_b\xi'''_b+6\xi''_a\xi''_b\xi'_b+\xi'^2_b\xi'''_a)+\cO(\epsilon^5),\\
b_{ab}(s)&=&\frac{1}{2}\epsilon^2(\xi''_a\xi'_b-\xi''_b\xi'_a)+\cO(\epsilon^4).
\eeqa
Note the even powers of $\epsilon$ in $\zeta$ and $b_{ab}$, and odd powers in $\mu_a$. This is due to the symmetry of the curve under $\epsilon\to -\epsilon$ and $\mathbf{n}^{(0)}_a\to-\mathbf{n}^{(0)}_a$ that maps $\mathbf{n}_a \rightarrow - \mathbf{n}_a$.
The conformal parameter $\s$ of the wavy line is related to $s$ by a reparametrization which can be expanded in terms of $\epsilon$
\begin{equation}
s(\s)=\s+\epsilon^2 s_2(\s)+\cO(\epsilon^4).
\end{equation}
The expansion starts at the second order in $\epsilon$ because $\zeta$ has no correction at first order.
Using the reparametrization properties \eqref{zetatrans}, \eqref{mutrans}, \eqref{btrans}, one can write down $\zeta(\s)$, $\mu_a(\s)$ and $b_{ab}(\s)$ in terms of $\xi_a$ and $s_{i}(\s)$
\begin{equation}\label{zetawavy}
\zeta(\s)=\epsilon^2 (2\dddot{s}_2+3\ddot{\xi}^2_a+2\dddot{\xi}_a\dot{\xi}_a)+\cO(\epsilon^4),
\end{equation}
\begin{equation}
\mu_a(\s)=\epsilon\dddot{\xi}_a-\frac{1}{2}\epsilon^3(4\dot{s}_2\dddot{\xi}_a+2s_2\ddddot{\xi}_a+\dot{\xi}_a\dot{\xi}_b\dddot{\xi}_b+6\ddot{\xi}_a\ddot{\xi}_b\dot{\xi}_b+\dot{\xi}^2_b\dddot{\xi}_a)+\cO(\epsilon^5),
\end{equation}
\begin{equation}
b_{ab}(\s)=\frac{1}{2}\epsilon^2(\ddot{\xi}_a\dot{\xi}_b-\ddot{\xi}_b\dot{\xi}_a)+\cO(\epsilon^4).
\end{equation}
Here we use dot to denote the derivative with respect to $\s$. One can solve the $\alpha$ in the Pohlmeyer reduction order by order with the boundary conditions $\mu_a$ and $b_{ab}$ and calculate $\zeta$ according to \eqref{Pb1}. Comparing it with \eqref{zetawavy} fixes the reparametrization up to the same order.

\subsection{Pohlmeyer equations}
The Pohlmeyer functions corresponding to the straight line are given by
\begin{equation}
\alpha=-\ln \tau,\quad u_a=0,\quad B_{ab}=0.
\end{equation}
For the wavy line, one can expand $\alpha$, $u_a$ and $B_{ab}$ around the straight line solution in terms of $\epsilon$:
\beqa
\alpha&=&-\ln \tau+\epsilon^2\alpha^{(2)}+\cO(\epsilon^4),\\
u_a&=&\epsilon u^{(1)}_a+\epsilon^3 u^{(3)}_a+\cO(\epsilon^5),\\
B_{ab}&=& \epsilon^2 B^{(2)}_{ab}+\cO(\epsilon^4),
\eeqa
The powers of $\epsilon$ are consistent with the boundary values. Plugging into \eqref{coshgordon}, we obtain the first few orders of the Pohlmeyer equations:
\beqa
\epsilon^1:\quad &\bar{\partial}u_a^{(1)}=0, \quad \partial \bar{u}_a^{(1)}=0,\label{ufirst}\\
\epsilon^2:\quad &4\partial\bar{\partial}\alpha^{(2)}-\frac{2}{\tau^2}\alpha^{(2)}=u^{(1)}_a\bar{u}^{(1)}_a\tau^2,\\
&\bar{\partial}B^{(2)}_{ab}-\partial\bar{B}^{(2)}_{ab}=\frac{1}{2}\tau^2(u^{(1)}_a\bar{u}^{(1)}_b-\bar{u}^{(1)}_au^{(1)}_b),\\
\epsilon^3:\quad&\bar{\partial}u_a^{(3)}=\bar{B}^{(2)}_{ab}u_b^{(1)}, \quad \partial \bar{u}_a^{(3)}=B^{(2)}_{ab}\bar{u}_b^{(1)}.
\eeqa
Continuing to higher orders, the equations should be solved recursively with the boundary conditions given in the previous subsection.

\subsection{Computing the area}
In this subsection, we compute the area for the wavy line to the leading order correction from the straight line and compare the result with the one obtained by Semenoff and Young in \cite{SY}.

To find the area, we need to perform the integral
\begin{equation}\label{areawavy1}
\mathcal{A}_f=-\epsilon^2\int d\sigma d\tau\tau^2 u^{(1)}_a\bar{u}^{(1)}_a,
\end{equation}
which requires solving for $u^{(1)}_a$. According to \eqref{ufirst}, $u^{(1)}_a$ is holormorphic and at the boundary one has
\begin{equation}
\Im(u^{(1)}_a(\sigma,\tau=0))=-\dddot{\xi}_a.
\end{equation}
At this order, there is no correction in the reparametrization, i.e., $s=\s$ and the functions $\xi_a(\s)$ describing the shape of the wavy line can be analytically continued to holomorphic functions $g_a(z)$ such that
\begin{equation}
\Re(g_a(\s,\tau=0))=\xi_a(\s).
\end{equation}
This leads to the identification
\begin{equation}
u^{(1)}_a(z)=-i\partial^3g_a(z).
\end{equation}
Therefore, \eqref{areawavy1} can be rewritten as
\begin{equation}\label{areaintegral1}
\mathcal{A}_f=-\epsilon^2\int d\sigma d\tau\tau^2 \partial^3g_a(z)\bar{\partial}^3\bar{g}_a(\bar{z}).
\end{equation}
Integrating by parts, \eqref{areaintegral1} reduces to
\begin{equation}
\begin{aligned}
\mathcal{A}_f&=-\epsilon^2\int d\s d\tau \partial(\tau^2 \partial^3 g_a \bar{\partial}^3 g_a+i \tau \partial g_a \bar{\partial}^3\bar{g}_a-\frac{1}{2}g_a\bar{\partial}^3\bar{g}_a)\\
&=-\frac{i}{2}\epsilon^2\int_{\mathbb{R}} d\bar{z} (\tau^2 \partial^3 g_a \bar{\partial}^3 g_a+i \tau \partial g_a \bar{\partial}^3\bar{g}_a-\frac{1}{2}g_a\bar{\partial}^3\bar{g}_a)\\
&=\frac{i}{8}\epsilon^2\int_{\mathbb{R}}d\s (\dddot{\bar{g}}_a g_a-\bar{g}_a\dddot{g}_a),
\end{aligned}
\end{equation}
where dots indicates $\partial_\sigma$. In the last equation we take $\tau=0$ and use that for a holomorphic function $\partial g_a(z)=\partial_\s g_a(z)$ and similarly for $\bar{g}(\bar{z})$. One can integrate by parts again and use some algebra to obtain
\begin{equation}\label{fareawavy}
\mathcal{A}_f
=\frac{1}{2}\int_{\mathbb{R}} d\s \ddot{\xi}_a\dot{\eta}_a
=-\frac{1}{2}\int_{\mathbb{R}} d\s \dot{\xi}_a\ddot{\eta}_a,
\end{equation}
where we define
\begin{equation}
\eta_a(\s)=\Im(g_a(\s,\tau=0)) .
\end{equation}
By the standard dispersion relation due to the analyticity of $g_{a}$, we have
\begin{equation}
\dot{\eta}_a(\s')=-\frac{1}{\pi} \fint d\s \frac{\dot{\xi}_a(\s)}{(\s-\s')}.
\end{equation}
where $\fint$ denotes principal value.
Plugging into \eqref{fareawavy}, integrating by parts and with some calculations, we find
\begin{equation}
\mathcal{A}_f=-\frac{1}{4\pi}\epsilon^2\int_{\mathbb{R}}ds ds'\frac{(\xi'_a(s)-\xi'_a(s'))^2}{(s-s')^2}.
\end{equation}
This agrees with the result obtained by Semenoff and Young in \cite{SY}.

\section{Conclusions}

In this work we studied the problem of finding a minimal area surface ending on an arbitrary curve in the boundary of \ads{n+1} by a non-trivial generalization of previous ideas applicable to \ads{3} \cite{WLMA,Dekel}. It is well-known that the equations of motion are integrable and can be greatly simplified using the Pohlmeyer reduction. In this paper we show that the boundary conditions for the Pohlmeyer reduction are given by conformal invariants of the curve \eqref{Pb1},\eqref{Pb2},\eqref{Pb3} but require choosing the correct conformal parameterization of the curve, namely the one that extends to conformal gauge in the world-sheet. Thus, we provide a test at the boundary, in terms of a further conformal invariant $\zeta(\s)$, that determines if we have the correct parameterization, namely the boundary equivalent of the Virasoro constraints in the world-sheet. 
 After that, by identifying a certain flat current \eqref{jArea} we obtained perhaps the most interesting result of the paper, a formula for the area in terms of generalized conformal arc-length, curvature and torsion \eqref{areaF}. These generalized invariants make use not only of the shape of the curve ($X^i_0(s)$ in \eqref{B5}) but also of a boundary condition given by the third normal derivative of the contour at the boundary ($X^i_3(\s)$ in \eqref{B6} or $\nu_a$  in \eqref{nudef}). This extra boundary condition can be related to the shape of the curve by solving the Pohlmeyer reduction problem and the self-consistency condition for the parameterization of the curve. From the point of view of integrability this relation between boundary conditions can be determined by imposing the vanishing of an infinite set of conserved quantities. Once the problem is set up, it can be solved numerically as in \cite{He:2017zsk} for \ads{3} or by an expansion near the straight-line solution as done in \cite{Dekel} for \ads{3} and here for \ads{n+1} (albeit only at the lowest order). If the techniques of \cite{Dekel} can be applied to this more general problem we can make contact of the four point functions recently computed in \cite{Giombi}. 

\section{Acknowledgments}

 We are very grateful to A. Dekel, N. Drukker, A. Tseytlin, and P. Vieira for useful comments and discussions on this topic. 
 This work was supported in part by DOE through grant DE-SC0007884. 

\appendix
\numberwithin{equation}{section}

\section{$\chi$, $Q$ and $T$ in $\mathbb{R}^3$ boundary of $\ads{4}$}\label{appA}

 In the case of three dimensional boundary $\mathbb{R}^3$, it is straightforward to find the expressions of conformal arc-length, conformal curvature and conformal torsion in terms of arc-length, curvature, and torsion of the curve. If we take the parameter $s$ to be arc-length, we have $|\mathbf{x}'|=1$ which simplifies the formulas of section \ref{boundarycurve}. Further, it is convenient to use the Frenet-Serre frame of normals $(\mathbf{n},\mathbf{k})$ defined as
\beq
 \mathbf{v}'= \kappa \mathbf{n}, \ \ \mathbf{n}'=-\kappa \mathbf{v} + \tau \mathbf{k}, \ \ \mathbf{k}'=-\tau \mathbf{n}.
\eeq
Here $\kappa$ is the ordinary curvature and $\tau$ the torsion. Using these formulas we find
\beqa
\zeta &=& \mathbf{v}'^2,\\
\mu_a &=& \mathbf{n}_a\cdot \mathbf{v}'', \\
D_s\mu_a &=& \mathbf{n}_a\cdot (\mathbf{v}'''+\kappa^3 \mathbf{n}),
\eeqa
leading to
\beqa
 \chi &=& \int_A^s \sqrt{\mu} ds =\int_A^s (\kappa'{}^2 + \kappa^2\tau^2)^{\frac{1}{4}} ds, \\
 Q &=& -\half\,\frac{\zeta-2\{\chi,s\}}{\mu} = \frac{1}{8\mu^3}(4\mu''\mu-4\kappa^2\mu^2-5\mu'{}^2), \\
T &=& \frac{|D\hat{\mu}_a|}{\sqrt{\mu}} = \frac{1}{\mu^{\frac{5}{2}}}(\kappa^2\tau^3-\kappa\kappa''\tau+\kappa\kappa'\tau'+2\kappa'{}^2\tau),
\eeqa
which agree with the standard definitions \cite{ConformalCurve}. Another formula for $\mu_a$ can be obtained by defining the curvatures
\beq
\kappa_a = \mathbf{n}_a \cdot \mathbf{v}', \ \ \ \ \ \Rightarrow \ \ \ \ \kappa_1 = \kappa,\ \kappa_2 =0,
\eeq
and then the simple formula
\beq
 \mu_a = D_s \kappa_a
\eeq
follows. Also, given a point on the curve, using symmetries, the curve in its neighborhood can be written as
\beqa
 x &=& x, \\
 y &=& \frac{x^3}{3!}+(2Q-T^2) \frac{x^5}{5!}+\cO(x^6), \\
 z &=& T \frac{x^4}{4!} + \frac{1}{\sqrt{\mu}} \frac{dT}{ds} \frac{x^5}{5!} + \cO(x^6).
\eeqa
As mentioned in the main text, the combination $2Q-T^2$ is the same as the one that appears in the formula for the area \eqref{areaF}, although there, it is in a generalized form where we replaced $\mu_a$ by $u_a$.

\section{Near boundary expansion in Poincare coordinates}\label{appB}
We consider the world-sheet to be the upper half plane with coordinates $(\sigma,\tau)$, the boundary being at $\tau=0$.  
The equations of motion are
\beqa
 (\partial_\s^2+\partial_{\tau}^2)X^i&=&\frac{2}{Z}\left(\partial_\s X^i \partial_\s Z+\partial_{\tau} X^i\partial_{\tau} Z\right), \\
 (\partial_\s^2+\partial_{\tau}^2)Z   &=&\frac{1}{Z}\left[(\partial_\s Z)^2+(\partial_\tau Z)^2-\partial_\s X^i\partial_\s X^i
 -\partial_\tau X^i\partial_\tau X^i \right].
\eeqa
 We expand the solutions near the boundary as
\beqa
 X^i(\sigma,\tau) &=& \sum_{n=0}^\infty X_n^i(\s) \tau^n, \\
 Z(\sigma,\tau)    &=& \sum_{n=1}^\infty Z_n(\sigma) \tau^n.
\eeqa
The first several coefficients of the expansion are \cite{Polyakov:2000jg}:
\beqa
X^i_0&=&\mbox{given boundary curve}, \  X_1^i=0, \  X_2^i=\half X''_0{}^i - \frac{Z'_1}{Z_1} X'_0{}^i, \label{B5}\\ 
X_3^i&=&\mbox{undetermined},\ \text{with} \ X_3\cdot X'_0=0, \label{B6}
\eeqa
\beq
Z_1 = |X'_0|, \  \  Z_2 =0,\ \  Z_3 = \frac{1}{3} \frac{Z'_1{}^2}{Z_1}-\frac{1}{6}Z_1''-\frac{1}{3}\frac{|X''_0|^2}{Z_1},
\eeq
where primes denote derivative with respect to $\s$. It is important to notice that $X_3^i(\s)$ is perpendicular to the tangent $X'_0{}^i$ but otherwise not determined. Finding $X_3^i$ requires solving the minimal area problem, namely imposing that the surface is regular. 
In \cite{Polyakov:2000jg} a very nice relation between the variation of the area and $X^i_3$ was given:
\beq
 X_3^i = -\frac{1}{3} |X'{}^i_0|^2\ \frac{\delta \cA_f}{\delta X^i_0}
\eeq
that shows $X^i_3$ behaves  like a momentum conjugate to $X^i_0$. Using this we can write, in the notation of section \ref{Pohlmeyerbc},
\beq
 \nu_a = - |x'{}^i_0|^2\ n^i_a \frac{\delta \cA_f}{\delta x^i_0}
\eeq 
as mentioned in the main text.
For higher order terms it is easy to derive a recursive relation that allows to obtain a high order expansion using a computer algebra program:
\beqa
X^{i}_{n+1} &=& \frac{1}{(n+1)(n-2)}\left[- X''^i_{n-1} + \frac{2Z'_1}{Z_1} X'^i_{n-1} +\frac{2}{Z_1}\sum_{p=0}^{n-2}(n-p+1)(p+1)Z_{n-p+1}X^i_{p+1} \right. \non \\
	&& \left. +\frac{1}{Z_1} \sum_{p=0}^{n-3}\left(-Z_{n-p}X''^i_{p}-(p+1)(p+2)Z_{n-p}X^i_{p+2}+2Z'_{n-p}X'^i_{p}\right)  \right], \\
Z_{n+1} &=& \frac{1}{n(n+1)}\frac{1}{Z_1} \left[-\sum_{p=1}^{n-1}Z_{n-p}Z''_{p}-\sum_{p=0}^{n-2}(p+1)(p+2)Z_{n-p}Z_{p+2}\right. \non \\
&& \left.+2\sum_{p=1}^{n-1}Z'_{n-p}Z'_p-2\sum_{p=1}^{n-1}(n-p+1)(p+1)X^i_{n-p+1}X^i_{p+1}\right].
\eeqa
 
 \section{AdS coordinates and $SO(n+1,1)$ generators}\label{appC}
 The space \ads{d=n+1} can be embedded into $\mathbb{R}^{d,1}$ as the subspace satisfying the constraint
 \beq
  Y^2 =\eta_{\mu\nu} Y^\mu Y^\nu = -1,
 \eeq
 with the metric
 \beq
  \eta = \mbox{diag}[1,\ldots,1,-1].
 \eeq
 It is also useful to solve the constraint in terms of  Poincare coordinates $(X_{i=1\ldots n},Z)$ as
 \beq
  Y_{i} = \frac{X_i}{Z}, \ \ Y_0=\frac{1}{2Z} (1+Z^2+X_i^2), \ \ Y_d=\frac{1}{2Z} (-1+Z^2+X_i^2).
 \eeq 
 We define the generators of $SO(n+1,1)$ to satisfy the commutation relations
 \beq
  [M_{\mu\nu},M_{\alpha\beta}] =- \eta_{\mu\alpha} M_{\nu\beta} + \eta_{\mu\beta} M_{\nu\alpha} + \eta_{\nu\alpha} M_{\mu\beta} - \eta_{\nu\beta} M_{\mu\alpha}.
 \eeq
 Two representations are particularly useful, the vector representation
 \beq
 \left(M^V_{\mu\nu}\right)_{\alpha}{}^{\beta} = \eta_{\mu\alpha}\delta_{\nu}^{\beta} - \eta_{\nu\alpha} \delta_{\mu}^{\beta},
 \eeq
 and the spinor representation
 \beq
  M^S_{\mu\nu} = \frac{1}{4} [\gamma_\mu,\gamma_\nu].
 \eeq
Throughout the paper we try to use final expressions that are independent of the representation  and rely only on the commutations relations.


\end{document}